\documentclass{article}

\usepackage{arxiv}

\usepackage[utf8]{inputenc} % allow utf-8 input
\usepackage[T1]{fontenc}    % use 8-bit T1 fonts
\usepackage{hyperref}       % hyperlinks
\usepackage{url}            % simple URL typesetting
\usepackage{booktabs}       % professional-quality tables
\usepackage{amsfonts}       % blackboard math symbols
\usepackage{nicefrac}       % compact symbols for 1/2, etc.
\usepackage{microtype}      % microtypography
\usepackage{lipsum}
\usepackage{verbatim} 
\usepackage{amsmath}
\usepackage{graphicx}
\usepackage{algorithm}
\usepackage[noend]{algpseudocode}

\title{\emph{A new approach on estimating the fluid temperature in a multiphase flow system using particle filter method}}

\author{
 Zhuoran Dang\thanks{zdang@purdue.edu} \\
 School of Nuclear Engineering\\
  Purdue University\\
  West Lafayette, IN 47907 \\
  }

\begin{document}
\maketitle

\begin{abstract}
Fluid temperature is important for the analysis of the heat transfers in thermal hydraulics. An accurate measurement or estimation of the fluid temperature in multiphase flows is challenging. This is due to that the thermocouple signal that mixes with temperature signals for each phase and non-negligible noises. This study provides a new approach to estimate the local fluid temperature in multiphase flows using experimental time-series temperature signal. The thermocouple signal is considered to be a sequence with Markov property and the particle filter method is utilized in the new method to extract the fluid temperature. A complete description of the new method is presented in this article. 
\end{abstract}


\section{Introduction}

Fluid temperature is fundamental to the analysis of heat transfers in heating and cooling systems. It is a parameter that can be experimentally measured or calculated and mainly used in the calculation of the heat transfer coefficients. Thus, accurate fluid temperature is very important for the estimation of heat transfer rates or coefficients. Heat transfer with phase change has been widely used in the current industrial applications such as boiling water reactors. This is because the heat transfer rate is larger in a flow system with phase change than one without phase change. However, the estimation of the heat transfer coefficient with phase change, namely, the interfacial heat transfer coefficient is harder than that of the signal phase heat transfer coefficient. This is mainly because the measurement of the fluid temperature in the multiphase flow system is challenging. The temperatures of each phase are different and the temperature signals are fluctuating. In this sense, the estimation of the fluid temperature in a multiphase flow system requires more studies. 

Despite it is a topic with several decade histories, the optimal approach of fluid temperature estimation is still not clear. Although some methods have been proposed, there is still no method that is recognized in consensus to be reliable. One simple and direct way to estimate the fluid temperature is to consider it to be equal to the lowest values in the temperature signals.\cite{warrier2002interfacial, jiji1964bubble} This is easy to process yet it lacks either theoretical or experimental validations. Besides, the lowest values are not the fluid temperature if there are non-negligible inherent noise in the data acquisition system. 
A more reliable and also more widely used method is to statistically extract the fluid temperature from the thermocouple signal by plotting the probability density function or histogram.\cite{delhaye1973void,roy1994local,beckman1993fast} In a steam-water flow system, for example, there are two characteristic peaks in the histogram and they correspond to the steam and water temperatures. This method is more reasonable and more robust than the above method. However, it highly relies on the quality of the thermocouple signal, that is, the thermocouple should be fast-response and signal-to-noise-ratio should be small. 
Recently, an algorithm that directly processes the thermocouple signal is developed \cite{kim2007measurement} that it discriminates the phases based on self-tuned thresholds and calculates the temperature of each phase based on the energy balance equation. This algorithm is more elaborated than the above two methods yet it still requires high-quality thermocouple signals.

The recently drastic development of computer science provides new solutions for every other subject related to it. In statistics and control theory, Kalman filtering, also known as linear quadratic estimation has been widely applied.\cite{kalman1960new} This algorithm uses time-series signals containing statistical noise and other inaccuracies and it estimates the unknown variable using a joint probability distribution over the variables for each time frame. It tends to be more accurate than those based on a single measurement alone. While the Kalman filter usually uses in a linear system with Gaussian distributed noises, Particle filter can extend the application range to the non-linear systems with non-Gaussian distributed noises.\cite{del1996non} The particle filter is a common application of the Sequential Monte Carlo method, which is a set of Monte Carlo algorithms used to solve filtering problems arising in signal processing and Bayesian statistical inference. The objective is to compute the posterior distributions of the states of some Markov process, given some noisy and partial observations. In this sense, it is possible to use the Particle filter for the estimation of fluid temperature using the thermocouple measurements in the multiphase flows. 

In this study, a new method for fluid temperature estimation based on particle filtering is designed. The theoretical basis of this method and a detailed description of this new method is presented. This new method is tested with time-series data generated using a microthermocouple measurements in subcooled boiling two-phase flow.

\section{Particle Filter}

The particle filter method is developed for the problem of a stochastic process with Markov property. This process contains both observable and hidden variables, and the observable variables are related to the hidden variables in some ways that can be described in known forms or models. In the hidden Markov model (HMM), the observable variables and hidden variables are used to describe the observation process and state process, respectively. The goal of the particle filter is to estimate the posterior state given the observable state.\cite{del1996non}

Define $X_{1:n}$ is the state of a Markov process and the observations $Z_{1:n}$  is the observation. The state $X_{t}$ is changed based on the transition probability function,
\begin{equation}
    X_{t}\left|X_{t-1}=x_{t} \sim p\left(x_{t} | x_{t-1}\right)\right.
\end{equation}
The observation $Z_{t}$ is assumed to be only related with the state $X_{t}$,
\begin{equation}
    Z_{t}\left|X_{t}=z_{t} \sim p\left(z_{t} | x_{t}\right)\right.
\end{equation}
The system can be modeled in the following state space equations:
\begin{equation}
    \begin{array}{l}{X_{t}=g\left(X_{t-1}\right)+W_{t-1}} \\ {Z_{t}=h\left(X_{t}\right)+V_{t}}\end{array}
\end{equation}
where $g$ and $f$ are both known functions. If $g$ and $f$ are linear and $W$ and $V$ are Guassian, then the system can be modeled with the Kalman filters to obtained the unbiased, optimal estimation. If not, we can design a particle filter if we could assume that we can create sample set for the transition of $X_{t}$ to $X_{t+1}$ and compute the weights/probabilities of the set.

Suppose that we are processing a time-series signal with a length of T. The number of the independent particle is N and the particles $X_{1:T}^{1:N}$ are independent of each other with density distributions of $p(X_{1:T}|Z_{1:T})$. $Z$ represents the observations. Note that each particle is a hypothesis about the current state.

For nonlinear filtering, recall that the conditional probability can be expressed with Bayes Rule: \cite{del1998measure}
\begin{equation}
    p\left(X_{t} | Z_{t}\right) = \frac{p\left(Z_{t} | X_{t}\right) p\left( X_{t}\right)}{p\left(Z_{t}\right)}
\end{equation}
where 
\begin{equation}
    \begin{array}{c}
    {p\left(Z_{t}\right)=\int p\left(Z_{t} | x_{t}\right) p\left(x_{t}\right) d x_{t}} \\ {p\left(Z_{t} | x_{t}\right)=\prod_{t=0}^{t} p\left(z_{t} | x_{t}\right)} \\ {p\left(x_{t}\right)=p_{0}\left(x_{0}\right) \prod_{t=1}^{t} p\left(x_{t} | x_{t-1}\right)}\end{array}
\end{equation}
Based on the above equations, the nonlinear particle filtering equation is given as follows, 
\begin{equation}
\begin{array}{c}
  {p\left(X_{t+1} |  Z_{t}\right)=\int p\left(x_{t+1} | x_{t}\right) p\left(x_{t} | Z_{t}\right) d x_{t}} \\
  {p\left(X_{t+1} | Z_{t+1}\right)=p\left(  Z_{t+1} | x_{t+1}\right) p\left(X_{t+1} |  Z_{t}\right)} \\
  {||p\left(X_{t+1} | Z_{t+1}\right)||=\frac{p\left(x_{t+1} | Z_{t+1}\right)}{\int p\left(Z_{t} | x_{t}^{\prime}\right) p\left(x_{t}^{\prime} | y_{t-1}\right) d x_{t}^{\prime}}}
  \end{array}
\end{equation}
The three equations in equation (6) corresponds to the evaluation, propagation, and normalization stage, which will be discussed in the later section. Besides, one of the important stage in the particle filtering that should be mentioned is resampling. Resampling allows us to obtain samples distributed approximately according to $p(x_{1:n})$.\cite{doucet2009tutorial} It removes particles with low weights and keeps the particles with a high weights. This is useful in a time-series modeling or sequential framework because the particles with low weights could be likely to turn into a high-weighted particle at the next time step and affects the estimation.\cite{doucet2009tutorial}
In the next section, the details of how to implement fluid temperature estimation using particle filtering are presented.

\section{Description of the new method}

The measurement of temperature using thermocouple is considered to be a stochastic process with Markov property. It means that the fluid temperature in the next state is assumed to be only related to the parameters in the current state. The fluid temperature estimation using the thermocouple signal can be regarded as a process of tracking with dynamics: given a model of expected fluid temperature in the current state, predict the value of fluid temperature in the next state. Therefore, the key idea of the particle filter is to generate a number of hypotheses, namely, particles, about the fluid temperature at time step t-1, $x_{t-1}$, and keep the most likely ones and propagate them further to $x_{t}$ with the measurement at time step t. Repeat these process again using  $x_{t}$ to $x_{t+1}$.
%%%

An important consideration is how to choose the model that describes the fluid temperature, which corresponds to the observation model. The mixed temperature in multiphase flow is simply modeled using a steady-state relation as follows,

\begin{equation}
    \begin{array}{c}
    {T_{\mathrm{m}}=\alpha_{1} T_{1}+\alpha_{2} T_{2}+\ldots+\alpha_{n} T_{\mathrm{n}}} \\ {\alpha_{1}+\alpha_{2}+\ldots+\alpha_{n}=1} \end{array}
\end{equation}

where $\alpha$ and $T$ are the local void fraction and temperature for each phase, respectively. In a two-phase flow system such as subcooled boiling flow, this equation reduces to a two-partition formulation and the steam temperature can be assumed to be equal to the saturation temperature. The mixed temperature is obtained from the thermocouple measurement and the local void fraction can be measured by various instrumentations such as conductivity probe. From the equation, the model calculates a constant value in a short measured period, which will definitely produce errors. However, the fluid temperature fluctuation is very complex to model even in a steady-state condition. This model approximates the fluid temperature and when it combines with the new method proposed in this study, the estimation accuracy can be improved.
%%%

\begin{figure}
     \centering
     \includegraphics[width=1\linewidth]{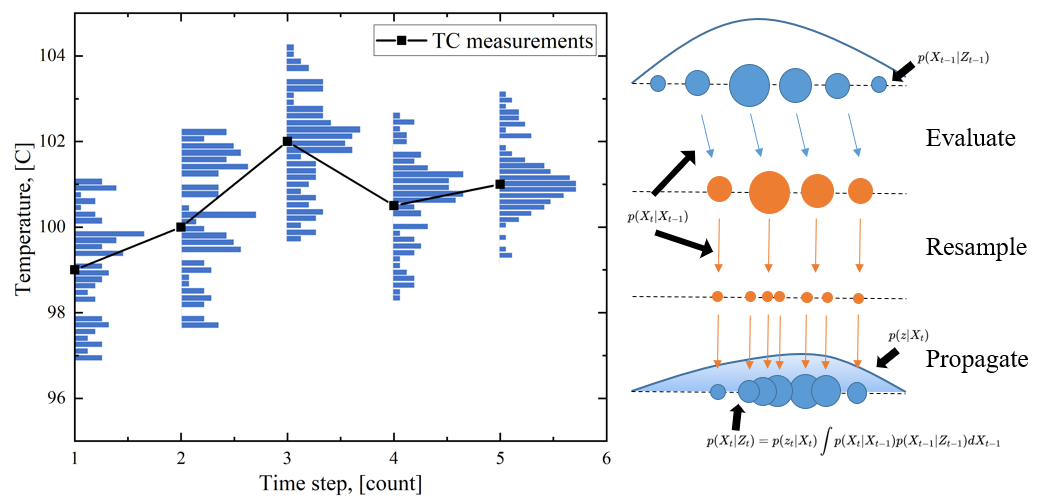}
     \caption{Conceptual diagram of fluid temperature estimate process with particle filter.}
     \label{fig1}
 \end{figure}

An example of this process is shown in the conceptual figure \ref{fig1}. The left part of the figure contains the average thermocouple measurement values marked in black dot and line in 5-time steps. The right part of the figure is a pipeline diagram that briefly explains the main procedures included for a single time step process. In the time step t-1, initially, the measured density distribution is represented by particle positions and weights, $p(X_{t-1}|Z_{t-1})$. The particles with relatively large weights are chosen and evaluated with the fluid temperature model, while the particles with small weights are discarded. This step is called as evaluation or prediction. In this process, the evaluation is to calculate the mean square error (MSE) between the particle value and the fluid temperature model. Then the evaluated particles are resampled based on their weights that particles with largely weights are represented by more particles, which will be evaluated in the next time step. In the propagation stage, some dynamics and random noise are added in each resampled particle. In this study, the dynamic model is simplified to a Gaussian distribution function with the center of the distribution equals to zero. In the final stage of measurement, the resampled particles are evaluated with the measurement density distribution at time step t. The algorithm is summarized below.
\begin{algorithm}[H]
\caption{}
\begin{algorithmic}
\State $\textit{initialize } x^{i}_{1} \sim p(x_{1}), w = \frac{1}{N}$.

\Loop{ for each time step t}
    \State $\textit{initialize } S_{t} = \emptyset, \eta=0 $.

\State $\textit{evaluate } w_{t}^{i}=p\left(x_{t}^{i} | x_{t}^{i-1}, u_{t} \right)\textit{ for i = 1 to N}$.
\State $\textit{resample and propagate} S_{t} = S_{t} \cup [x_{t}^{i}]_{i=1}^{N} \textit{ from } [x_{t}^{i}, w_{t}^{i}]_{i=1}^{N}$.
\State $\textit{compute weights with measurement }w_{t}^{i}=p\left(z_{t} | x_{t}^{i}\right), \eta=\eta+w_{t}^{i}$.
\State $\textit{normalize weights } w_{t}^{i}=w_{t}^{i} / \eta $.
\EndLoop
\State $\textbf{end}$
\end{algorithmic}
\end{algorithm}

\section{Experiment}

The method is tested with the time-series signals measured using a microthermocouple in a steady-state, subcooled boiling, two-phase flow condition. The local void fraction is measured using a conductivity probe sensor near the same location. The fluid temperature is estimated using this time-averaged void fraction and mixed fluid temperature, as presented above. 

\begin{figure}
     \centering
     \includegraphics[width=1\linewidth]{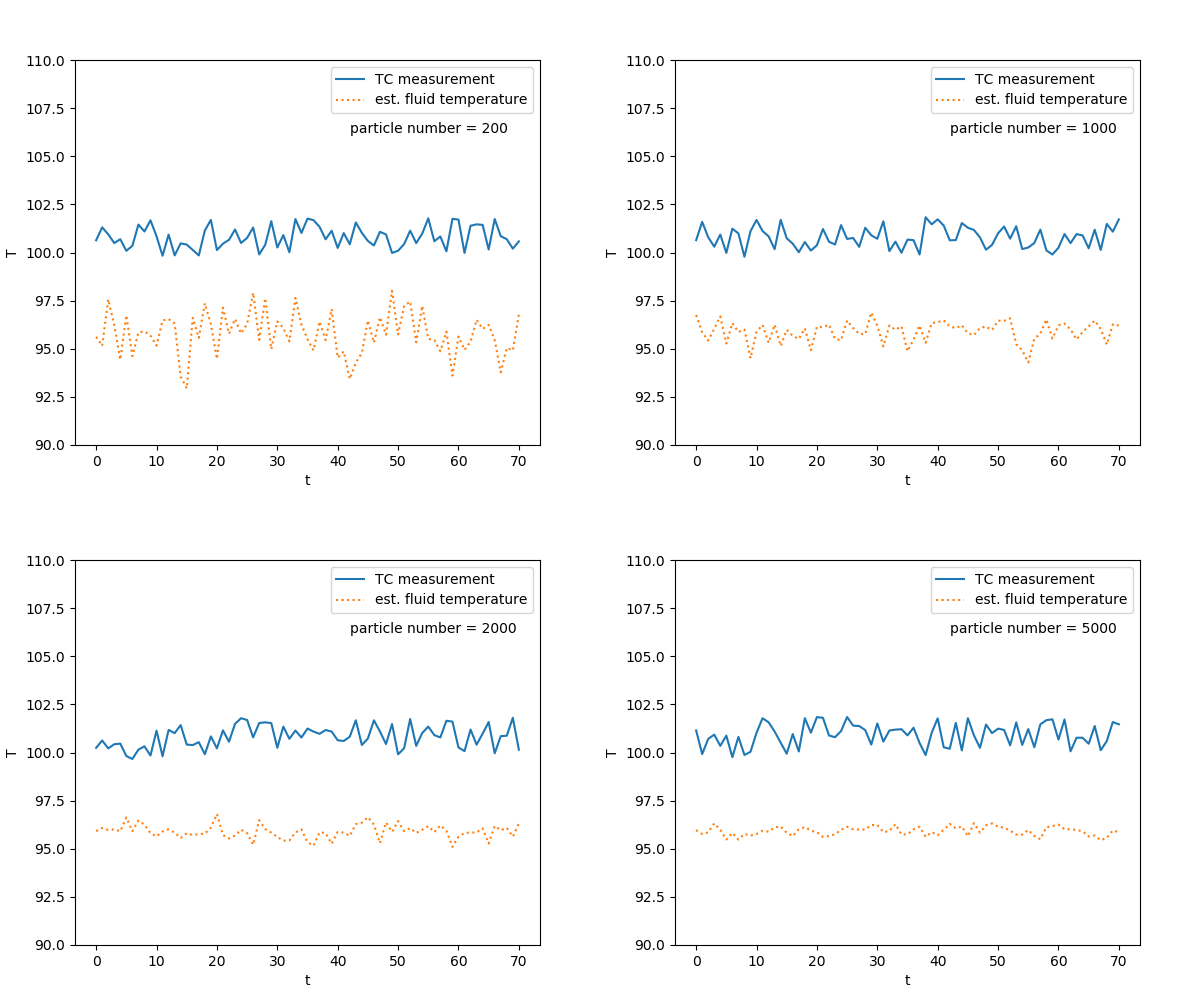}
     \caption{Fluid temperature estimation with different numbers of particles.}
     \label{fig2}
 \end{figure}

The performance of the particle filters is largely related with the observation model, dynamic model, and the number of the particles.\cite{} The observation model and the dynamic model varies from different modeling subject and setup, therefore, analyzing their effect on the model performance may not be worthwhile. The effect of the number of particles is analyzed in this study, which is shown in the results figures \ref{fig2}. It can be easily seen that with a larger number of particles, the oscillation of the estimation curve is suppressed. However, the effect on the accuracy of the estimation is not clearly observed when comparing the results, yet it is reported in many studies. This may due to that the changes in the thermocouple signal are not significant.

\section{Conclusion}
This study provides a new method to estimate the local fluid temperature in multiphase flows using experimental time-series temperature signal. This new method is designed based on the particle filter algorithm and it assumes that the thermocouple signal is with Markov property, that is, the future state of the temperature conditioned on both the past and present states depends only on the present state. \cite{gudivada2015big} In this method, a batch of particles/hypothesis about the fluid temperature is generated and evaluated. The particles with a high probability or weight are kept and propagate to the next time step. This method is tested with the time-series temperature signal generated with a microthermocouple in a subcooled boiling two-phase flows and it is observed that the oscillation of the estimation curve is strongly related with the number of particles.

\section{Acknowledgements}
The author is currently a Ph.D. student in thermal hydraulics and reactor safety laboratory (TRSL) at Purdue University and under the supervision of Dr. Mamoru Ishii, Walter Zinn Distinguished Professor of Nuclear Engineering. The author would like to deeply thank his support and guidance in the theory of thermo-fluid dynamics and two-phase flow. 

\bibliographystyle{unsrt}  
\bibliography{references}

\end{document}